\documentclass[sigconf]{acmart}

\AtBeginDocument{%
  \providecommand\BibTeX{{%
    \normalfont B\kern-0.5em{\scshape i\kern-0.25em b}\kern-0.8em\TeX}}}

\copyrightyear{2022}
\acmYear{2022}
\setcopyright{acmlicensed}
\acmConference[GECCO '22 Companion]{Genetic and Evolutionary Computation Conference Companion}{July 9--13, 2022}{Boston, MA, USA}
\acmBooktitle{Genetic and Evolutionary Computation Conference Companion (GECCO '22 Companion), July 9--13, 2022, Boston, MA, USA}
\acmPrice{15.00}
\acmDOI{10.1145/3520304.3533952}
\acmISBN{978-1-4503-9268-6/22/07}

\usepackage{graphicx}
\usepackage{algorithm2e}
\usepackage{algorithmic}
\usepackage{amsmath}
\usepackage{amsfonts}
\usepackage{hyperref}

\begin{document}

\title{Algorithmic QUBO Formulations for $k$-SAT and Hamiltonian~Cycles}


\author{Jonas Nüßlein}
\email{jonas.nuesslein@ifi.lmu.de}
\affiliation{%
  \institution{LMU Munich}
  \country{Germany}
}

\author{Thomas Gabor}
\email{thomas.gabor@ifi.lmu.de}
\affiliation{%
  \institution{LMU Munich}
  \country{Germany}
}

\author{Claudia Linnhoff-Popien}
\email{linnhoff@ifi.lmu.de}
\affiliation{%
  \institution{LMU Munich}
  \country{Germany}
}

\author{Sebastian Feld}
\email{s.feld@tudelft.nl}
\affiliation{%
  \institution{TU Delft}
  \country{Netherlands}
}

\renewcommand{\shortauthors}{Nüßlein et al.}


\begin{abstract}
Quadratic unconstrained binary optimization (QUBO) can be seen as a generic language for optimization problems. QUBOs attract particular attention since they can be solved with quantum hardware, like quantum annealers or quantum gate computers running QAOA. In this paper, we present two novel QUBO formulations for $k$-SAT and Hamiltonian Cycles that scale significantly better than existing approaches. For $k$-SAT we reduce the growth of the QUBO matrix from $O(k)$ to $O(log(k))$. For Hamiltonian Cycles the matrix no longer grows quadratically in the number of nodes, as currently, but linearly in the number of edges and logarithmically in the number of nodes.

We present these two formulations not as mathematical expressions, as most QUBO formulations are, but as meta-algorithms that facilitate the design of more complex QUBO formulations and allow easy reuse in larger and more complex QUBO formulations.
\end{abstract}

\keywords{QUBO, Ising, Satisfiability, $k$-SAT, Hamiltonian Cycle}

\maketitle

\RestyleAlgo{ruled}

\section{Introduction}


Solving optimization and decision problems is a central task in computer science with numerous real-world applications~\cite{1,2,3}. However, not all problems can be solved exactly in polynomial time, assuming that the conjecture $P \neq NP$ \cite{cook2000p} holds.

The complexity class $P$ contains the problems which can be solved exactly with a polynomial-time algorithm.
In $NP$ are the problems for which a given solution can be verified in polynomial time, whereby $P \subseteq NP$.
Another complexity class worth mentioning is $NP$-hard, which consists of the problems for which probably no polynomial solution algorithm exists. A problem is called NP-complete if it is in both $NP$ and $NP$-hard.

However, numerous relevant problems, such as the Traveling Salesman Problem (TSP) or Satisfiabiliy (SAT), lie in $NP$-hard ~\cite{1,2,3,impagliazzo2001complexity}. For this reason, among others, there has been a recent growth of interest in quantum computers, with the hope that their non-deterministic computations will provide advantages in solving optimization and decision problems.
\ \\

There are two basic approaches to quantum computing, the Quantum Gate Model \cite{zahedinejad2017combinatorial, shor1999polynomial, grover1996fast} and Adiabatic Quantum Computing \cite{albash2018adiabatic, kaminsky2004scalable}, with Quantum Annealing \cite{finnila1994quantum, mcgeoch2014adiabatic} in particular. While Quantum Annealing is specifically designed to solve optimization problems, Gate Model Quantum Computing is somewhat more general, with the Quantum Approximate Optimization Algorithm (QAOA) \cite{farhi2014quantum} existing here, which can solve optimization problems. Since current quantum computers are still small and error-prone \cite{proctor2022measuring}, recent work has also been done on quantum-inspired classical computers, such as the Digital Annealer \cite{aramon2019physics}, which is also specifically designed to solve optimization problems. All of the solution methods mentioned here use Quadratic Unconstrained Binary Optimization (QUBO) \cite{glover2018tutorial} or isomorphic Ising \cite{barahona1982computational} as the formulation language for the optimization problems to be solved.

Thus, in order to solve an optimization problem using the above methods, it must first be translated to QUBO. The translation should be as efficient as possible in order to be able to solve larger problem instances on the limited hardware.
\ \\

In this paper, we would like to contribute by presenting two QUBO formulations for (Max) $k$-SAT and Hamiltonian Cycles which scale better than the currently existing ones. We further present the QUBO formulations not as mathematical expressions, but as algorithmic functions, making them easier to understand and also easier to reuse, for example as part of bigger and more complex QUBO formulations.
\ \\

\section{Background}

In this section, we want to formally introduce the problem classes QUBO, (Max) SAT, and Hamiltonian Cycles.

\subsection{Quadratic Unconstrained Binary Optimization (QUBO)}

Given a symmetric (n$\times$n)-matrix $Q$ and a binary vector $x$ of length $n$, a QUBO ~\cite{10} is a function of the form:

\begin{equation}
H(x,Q)=\sum_{i=1}^{n}\sum_{j=i}^{n}{x_i\ x_j\ Q_{ij}}
\end{equation}

The function $H$ is called Hamiltonian.
We will refer to the matrix $Q$ as the ``QUBO matrix'' in this paper.

The optimization task is to find a binary vector $x$ which is as close to the optimum $x^{\ast}=argmin_x \; H(x,Q)$ as possible. This we want to delegate to the machine. Our task, on the other hand, is to specify a function for a problem class such as TSP, which maps a concrete problem instance $P$ from this problem class to a QUBO matrix $Q$ in such a way that the solution $p$ (e.g. the shortest route in TSP) for the problem instance $P$ can be derived from the solution vector $x^{\ast}$.

Numerous well-known optimization problems such as boolean formula satisfiability (SAT), knapsack, graph coloring, the traveling salesman problem (TSP), or max clique have already been translated to QUBO form~\cite{8,12,13,15,16}. Throughout literature these translations have mostly been given via purely arithmetic expressions without imperative control structures~\cite{7,12,13,15,16}.

To solve a QUBO matrix using quantum annealing (QA), it must first be embedded on a special graph ~\cite{4,10,12}, where the nodes represent the qubits and the edges represent the connections. In this paper, we will not elaborate on embedding the QUBO matrices of our algorithms on these graphs.

\subsection{Satisfiability (SAT)}

Satisfiability (SAT) is one of the best known and most fundamental problems in computer science. In SAT a set of boolean variables $X$ is given, as well as a boolean formula $f$, which contains only variables from $X$. The question to be answered is whether there is an assignment $\underline{X}$ for the variables $X$ with truth values (1 and 0), so that $f$ evaluates to $1$. If there is such an assignment $\underline{X}$, it is called model for $f$ and $f$ is called satisfiable.

Any boolean function can be written in conjunctive normal form. Conjunctive normal form is a conjunction over any number of clauses, where a clause $C$ is a disjunction over any number of literals (a literal $l$ is a variable $x$ or its negation $\lnot x$). Assuming there are $|f|$ clauses, then $f$ can be written as
\begin{equation}
f=\bigwedge_{i=1}^{|f|}{\bigvee_{l \in C_i} l}
\end{equation}

Thus, for $f$ to evaluate to 1, each clause $(\bigvee_{l \in C_i}{l})$ must evaluate to 1. $k$-SAT is a special case of SAT where each clause contains exactly $k$ literals. For $k\geq3$ $k$-SAT is NP-complete~\cite{20}, for 2-SAT there exists a polynomial algorithm~\cite{21}. Max $k$-SAT is the optimization problem where the aim is to find the assignment that satisfies as many clauses as possible. If $f$ is satisfiable, then every solution for Max $k$-SAT is also a model for $k$-SAT. Algorithms for Max $k$-SAT therefore trivially also solve $k$-SAT.

\subsection{Hamiltonian Cycle}

In graph theory a cycle is a path following the edges of that graph where only the start and the end vertices are equal. In the Hamiltonian Cycle problem, a graph with a set of vertices $V$ and a set of directed or undirected edges $E$ is given. The question now is whether there is a cycle that visits every vertex of the graph. For each consecutive pair of vertices $(a,b)$ in the cycle there must be a corresponding (directed) edge in $E$. The Hamiltonian Cycle problem is NP-complete as well~\cite{22}. The Hamiltonian Cycle problem is a special case of TSP, where all edge weights are equal.

\section{Related Work}
\label{sec:rel}

Chancellor et al.~\cite{14} presented a QUBO formulation for Max $k$-SAT in which each clause is represented by Hamiltonian $H_{clause}^{(2)}$. For a clause with $k$ literals $(\sigma_i^z)$, $k$ ancillae $(\sigma_{i,a}^z)$ are necessary. Each clause Hamiltonian is given via 
\begin{multline}
H_{clause}^{(2)}=J\sum_{i=1}^{k}\sum_{j=1}^{i-1}{c(i)c(j)}\sigma_i^z\sigma_j^z + h\sum_{i=1}^{k}{c(i)\sigma_i^z} + \\ J^a\sum_{i=1}^{k}\sum_{j=1}^{i-1}{c(i)\sigma_i^z\sigma_{j,a}^z}+\sum_{i=1}^{k}h_i^a\sigma_{i,a}^z
\end{multline}

where $c(i)=1$ if literal $i$ is positive (not a negation) and $c(i)=-1$ if literal $i$ is a negation. The parameters $J$, $h$, and $h_i^a$ are chosen as $J=J^a$, $h=-J^a$, $h_i^a=-J^a(2i-k)+q_i$ with \par
\ \\
$q_i = \begin{cases}
g/2 & \text{ if } i=0 \\
0 & \, \text{ else,}
\end{cases}$ \par
\ \\
where $g/2 \ll J_a$. The Hamiltonian of the entire formula is then obtained by adding all the clause Hamiltonians.

Choi~\cite{8,13} used a different approach for 3-SAT, which can be easily generalized to $k$-SAT and also grows linearly in $k$.
\ \\

Lucas~\cite{16} presented the current state-of-the-art QUBO formulation for Hamiltonian Cycle, which grows quadratically in the number of vertices $N$. The QUBO matrix is given via
\begin{multline}
H=A\sum_{v=1}^{n}\left(1-\sum_{j=1}^{N}x_{v,j}\right)^2+A\sum_{j=1}^{n}\left(1-\sum_{v=1}^{N}x_{v,j}\right)^2 + \\
A\sum_{\left(uv\right)\notin E}\sum_{j=1}^{N}{x_{u,j}x_{v,j+1}}
\end{multline}

The first summand represents the constraint that each node appears at exactly one position of the cycle, the second summand represents the constraint that there is only one node at each position of the cycle, and the third summand ensures that two adjacent nodes in the cycle really have an edge connecting them. 

Vargas-Calderón et al.~\cite{23} presented a TSP formulation (Hamiltonian Cycle is TSP with equal edge weights), which requires only $N \cdot \log(N)$ qudits, where N-level qudits must be available. Our approach, however, requires only qubits (2-level qudits).

Besides quantum annealing, there are also gate model approaches to solve the Hamiltonian Cycle problem~\cite{26,27}, where~\cite{26} is an explicit algorithm for Hamiltonian Cycles, while~\cite{27} proposes the Grover algorithm~\cite{29} to find the Hamiltonian Cycle, whereby Grover provides quadratic speedup compared to brute force.
\ \\

Although there has been an approach~\cite{30} to code the constrains of an optimization problem with a programming language, which allows for loops and control structures which is then translated via hard-wired mechanisms into a QUBO matrix, still most QUBO formulations are presented via a mathematical expression as we have seen above. Our approach (algorithmic QUBO formulation) goes in a different direction: We formulate the QUBO not directly (neither via a mathematical expression nor via classical code which can be directly translated to a QUBO matrix) but indirectly via classical code which \emph{outputs} a QUBO. If we want to consider the QUBO matrix as the ``program'' for a quantum annealer, then the classical program thus acts as a meta-program.
\ \\

In this paper we want to show why algorithmic QUBO formulations are superior to mathematical QUBO formulations since mathematical QUBO formulations get complicated and chaotic very quickly as the complexity of the underlying constraints increases. We present algorithmic QUBO formulations for (Max) k-SAT and Hamiltonian Cycles for which we make use of the logarithm trick, which was already mentioned in ~\cite{12} and ~\cite{16}.

\section{Algorithmic QUBO Formulation for (Max) $k$-SAT}

As mentioned in Section~\ref{sec:rel}, the QUBO matrix grows linearly in $k$ in the current state-of-the-art formulation for (Max) $k$-SAT. We now show an algorithmic QUBO formulation for (Max) $k$-SAT which grows only logarithmically in $k$. We first present the basic idea behind our algorithm, followed by the more detailed description of the algorithm, an analysis of the scaling function of the QUBO matrix as a function of $k$, and a discussion.

\subsection{Idea}

Given a clause $C$, consisting of $k$ literals. The idea is to model a linear equation which counts by how many literals $C$ is satisfied. To count from 0 to $k$ we need $h=\left\lceil\log_2{(k+1)}\right\rceil$ binary variables $({A}_1,\ ... ,\ {A}_h)$, which we call auxiliary variables. If the clause is satisfied by at least one literal, at least one auxiliary will have value 1. We can thus form an artificial clause $newC = [{A}_1\ \vee\ \ ..\ \ \vee\ {A}_h]$, which will evaluate to true if $C$ evaluates to true and which evaluates to false if $C$ evaluates to false. We proceed recursively with anchor 2-SAT and 3-SAT for which we use the known formulations ~\cite{14,15}. Since the formulations for 2-SAT and 3-SAT actually solve Max 2-SAT and Max 3-SAT, our algorithm actually also solves Max $k$-SAT, which trivially solves $k$-SAT.

\subsection{Algorithm}

We first present the linear equation which counts by how many literals a clause is satisfied.

$n(C)$ is the number of negative literals in $C$ and $x(l)$ returns the assignment ($0$ or $1$) of the variable of literal $l$. $sign(l)$ returns $-1$ if $l$ is a negation and $+1$ if it is not a negation. The linear equation is thus given by formula (5)

\begin{equation}
n(C)+\sum_{l \in C}{sign\left(l\right) \cdot x\left(l\right)}=\sum_{j=1}^{h}2^{j-1} \cdot {x\left(A_j\right)}
\end{equation}

The auxiliary variables (right side of the equation) model the number of literals that satisfy clause $C$. If all $h$ auxiliary variables are $0$, then clause $C$ is not satisfied. For $C$ to be satisfied, at least one literal must satisfy the clause and thus the right-hand side of the equation must be greater than $0$, i.e., at least one of the $h$ auxiliary variables must be 1.

Ensuring that at least one of the auxiliary variables is $1$ corresponds to another clause $[{A}_1\ \vee\ \ ...\ \vee\ {A}_h]$, which only has $h$ (positive) literals.

We proceed in this way recursively until the clause consists of two or three literals, for which we then use the well-known formulations for OR (for clauses with two literals) or the 3-SAT matrices from~\cite{14,15} as anchors of the recursion.

In order to integrate the linear equation into a QUBO, it must be represented as a quadratic optimization function. This is easily done by putting all terms of the equation on one side and then squaring it (see formula (6)):
\begin{equation}
{\left( n(C) + \sum_{l \in C}{sign\left(l\right) \cdot x\left(l\right)} - \sum_{j=1}^{h}2^{j-1} \cdot {x\left(A_j\right)} \right)}^2
\end{equation}
\ \\

Let $f$ be the set of clauses, $X$ the set of variables, and $n(C)$ the number of negative literals in the clause $C$. $C[i]$ is the $i$-th literal of clause $C$, $x(C[i])$ is therefore the assignment ($0$ or $1$) of the corresponding variable of that literal. $x(A_j)$ is correspondingly the assignment of the $j$-th auxiliary variable. $len(C)$ returns the number of literals in clause $C$. Algorithm 1 shows the computation of the QUBO matrix $Q$.

\begin{algorithm}
  \label{alg:maxksat}
  \caption{QUBO algorithm for (Max) $k$-SAT}
  \SetKwFunction{algo}{fillQ}
  \SetKwFunction{proc}{implementClause}
  
  \SetKwProg{myalg}{}{}{}
  \myalg{\algo{Formula: $f$}}{
  \nl init empty QUBO matrix $Q$\;
  \nl \For{$C \in f$}{
      \nl \proc{$Q$, $C$}\;
  }
  \nl \KwRet $Q$\;}{}
  
  \ \\
  \ \\
  
  \SetKwProg{myproc}{}{}{}
  \myproc{\proc{QUBO matrix: $Q$, Clause: $C$}}{
  
  \nl \uIf{$len(C) = 2$}{
    \nl $Q \leftarrow OR(C)$\;
  }
  \nl \uElseIf{$len(C) = 3$}{
    \nl $X \leftarrow [A_1]$\;
    \nl $Q \leftarrow$ 3-SAT$(C,A_1)$\;
  }
  \nl \Else{
    \nl $h=\left\lceil\log_2{(len(C)+1)}\right\rceil$\;
    \nl $X \gets [{A}_1\ ,\ \ ...\ \ ,\ {A}_h] $\;
    \nl $Q \leftarrow$ formula (6)\;
    \nl $newC = [{A}_1\ \vee\ \ ..\ \ \vee\ {A}_h]$\;
    \nl \proc{$Q$, $newC$}\;
  }
  
  \nl \KwRet $Q$\;}
  
\end{algorithm}

``$Q\gets \textit{Formula}$'' can be interpreted as the addition of $\textit{Formula}$ to the current matrix $Q$ (simple element-wise addition). Note that different auxiliary variables are needed for every clause. Thus, in lines 8 and 12 of the algorithm, the size of $Q$ grows by 1 and $h$, respectively.

If $f$ is satisfiable, then the assignment that satisfies $f$ is encoded in the first $|X|$ elements of the solution vector $x^\ast$. The QUBO matrix built from this algorithm only grows logarithmically in $k$ and it needs additionally one QUBO variable for each SAT variable. The algorithm uses imperative statements and control structures as well as recursion. Besides that, it reuses the QUBO formulation of the OR function, the 3-SAT function, and recursively the function itself. Once a better way to encode 3-SAT is found, we only have to plug in the new code in lines 8 and 9 of our algorithm, without any further changes.

\subsection{Size of $Q$ as a function of $k$}

Assuming the original boolean formula $f$ has $|X|$ variables and $|f|$ clauses, the size of the QUBO matrix now no longer grows linearly in $k$, as in the current state-of-the-art, but logarithmically.
\ \\

The size of the resulting QUBO matrix $Q$ (the quadratic $(n \times n)$-matrix) can be recursively calculated via  $n=|X|+|f|\ \cdot \ r(k)$, where 
$$r(k) = \begin{cases}
0 & \text{if } k=2, \\
1 & \text{if } k=3,\\
\left\lceil\log_2{(k+1)}\right\rceil+r(\left\lceil\log_2{(k+1)}\right\rceil)\ & \text{if } k \geq 4.
\end{cases}$$
\ \\

\subsection{Discussion}


As in~\cite{14}, we also formulate each clause individually and the QUBO matrix for the whole formula is then obtained as the (element-wise) addition of all clause QUBO matrices. Figure 1 plots the scaling behaviors of the previous state-of-the-art QUBO formulation and our QUBO algorithm for $k$-SAT. The x-axis of this graph represents the number of literals in the clause $k$ and the y-axis represents the number of auxiliary variables needed. As can be seen, for $k=8$ our algorithm still needs as many auxiliary variables as~\cite{14}, i.e., $8$ auxiliary variables ($4+3+1$). Thus, our QUBO algorithm for $k$-SAT is really advantageous only for $k\gg8$, which is not feasible on current quantum computers.

Therefore, to verify the correctness of our QUBO algorithm, we solved QUBO matrices using a classical QUBO solving method, more specifically QbSolv~\cite{boothpartitioning}. We created 30 random, solvable formulas for $k=4,6,8,10$ for different clause-to-variable ratios. By ``solvable'' we mean that we used a classical SAT solver (Minisat~\cite{een2003extensible}) to check whether a model exists for the formula (i.e., whether it is satisfiable). Then we solved the QUBO matrices using QbSolv. For all formulas, the best QUBO solution corresponded to a model of the formula, which means that the QUBO formulation was indeed correct for these 120 formulas.

\begin{figure}[tb]
 \centering
 \includegraphics[scale=0.55]{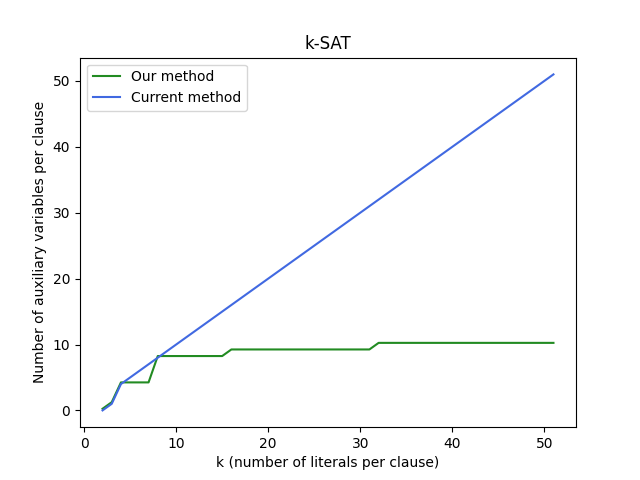}
 \caption{The number of necessary auxiliary variables per clause as a function of the number of literals in that clause for the current state-of-the-art implementation for $k$-SAT~\cite{14} compared to our QUBO algorithm for $k$-SAT.}
\end{figure}

\section{Algorithmic QUBO Formulation for Hamiltonian Cycle}

In this section, we consider the Hamiltonian circle problem. Without loss of generality we assume all edges to be directed. We use the logarithm trick ~\cite{12,16} and present an algorithmic QUBO formulation which only grows linearly in the number of edges and logarithmically in the number of nodes of the graph, which together can be sub-quadratic given that not all possible edges are used.

\subsection{Idea}

The current SOTA formulation for Hamiltonian Cylce formulates the optimization problem as a search for which of the $n-1$ vertices is positioned at which of the $n-1$ positions of the cycle.

We propose a different approach which is based on the search for the edges which together constitute the Hamiltonian Cycle. Thus, a solution $x$ consists of the position numbers in the cycle for all directed edges of the graph. Each position number is represented by $z=\left\lceil{log}_2(|V|+1)\right\rceil$ binary variables of $x$.

$P_{ab} \in \mathbb{N}_0$ is the position of the edge $(a,b)$, which is given by the $z$ position variables $x_{(a,b),1}, ..., x_{(a,b),z}$ and the linear equation

\begin{equation}
P_{ab}=\sum_{i=1}^{z}{2^{(i-1)}\ \cdot\ x_{(a,b),i}}
\end{equation}

Here, $x_{(a,b),i}$ is the assignment value (0 or 1) of the $i$-th variable of the subgroup of the QUBO solution vector $x$ belonging to the edge $(a,b)$. If the edge $(a,b)$ is not part of the cycle, then by definition $P_{ab}=0$.
\ \\

The Hamiltonian Cycle problem is fully described by the following three hard-constraints:
\begin{enumerate}
    \item The continuation of the cycle is deterministic, i.e., for all edge pairs $(a,b)$ and $(c,d)$ in the Hamiltonian Cycle $a \neq c$ and $b \neq d$ holds.
    \item Every edge has a continuation, i.e., if the edge $(a,b)$ is part of the cycle, there exists an edge $(b,c)$ for some $c$ in the cycle.
    \item There is an edge out of the start node $V_1$ (an arbitrary node of the graph) at position 1 and an edge back into the start node at position $|V|$, where $|V|$ is the number of vertices of the graph.
\end{enumerate}

Constraint (1) is very easy to implement by preventing the simultaneous activation of two variables from the subgroups of two conflicting edges by an appropriately high value in the matrix $Q$. The value $2{|V|}^2$ is sufficiently high. Constraint (2) can be ensured by the following objective: $Q\ \gets\left(P_{bc}-P_{ab}-1\right)^2$. This objective ensures that if the edges $(a,b)$ and $(b,c)$ are part of the Hamiltonian Cycle, then for the position $P_{bc}$ it holds that $P_{bc}=P_{ab}+1$. Constraint (3) actually allows us to reduce the number of needed variables further: Since edges of the form $(V_1,a)$ must be either at position $0$ (i.e., not part of the Hamiltonian Cycle) or at position $1$, only $z=1$ variables are needed for these edges. The same holds for edges of the form $(z,V_1)$, which must be either at position $0$ or $|V|$. The objectives that model these constraints as a quadratic function are given by
\begin{equation*}
Q\ \gets\left(P_{V_1a}-1\right)^2 \ \ \ \ \ \ \text{and}\ \ \ \ \ \ 
Q\ \gets\left(P_{zV_1}-|V|\right)^2.
\end{equation*}

\subsection{Algorithm}

To summarize the idea, an edge $(b,c)$ which has no contact with the starting node (i.e., $b \neq V_1$ and $c \neq V_1$) must satisfy the two linear equations $P_{bc}=P_{ab}+1$ and $P_{cd}=P_{bc}+1$. The addition of the two associated quadratic optimization problems $\left(P_{bc}-P_{ab}-1\right)^2$ and $\left(P_{cd}-P_{bc}-1\right)^2$ results in:  $\left(-2P_{bc}P_{ab}-2P_{cd}P_{bc}\right)+2{P_{bc}}^2+\left({P_{ab}}^2+2P_{ab}\right)+\left({P_{cd}}^2-2P_{cd}\right)$. As one can see, the part concerning only the edge $(b,c)$ is given by $2{P_{bc}}^2$. Plugging in the definition of $P_{bc}$ results in $2\left(\sum_{i=1}^{z}{2^{(i-1)}\ \cdot\ x_{(b,c),i}}\right)^2$. This constraint is now in a usual quadratic form, which can be added (element-wise addition) to the QUBO matrix. \par
If the edge is an edge from the start node, i.e., of the form $(V_1,a)$, then the sum of the two objectives $\left(P_{V_1a}-1\right)^2$ and $\left(P_{ab}-P_{V_1a}-1\right)^2$ results in $2{P_{V_1a}}^2$ as the part which only concerns the edge $(V_1,a)$. If it is an edge into the start node, i.e., of the form $(z,V_1)$, then the entries for the QUBO matrix $Q$ are the sum of $\left(P_{zV_1}-\left|V\right|\right)^2$ and $\left(P_{zV_1}-P_{yz}-1\right)^2$. Expanded, the part concerning only the edge $(z,V_1)$ is equal to $2{P_{zV_1}}^2-2P_{zV_1}\cdot(\left|V\right|+1)$. Putting it all together, the QUBO is generated by Algorithm 2.
\ \\

Note that only a few of the above constraints are actually satisfied by the correct solution $x^\ast$. For example, given there is an edge $(a,b)$ and three possible continuations $(b,c)$, $(b,d)$ and $(b,e)$. For each edge pair there is a corresponding optimization problem: ${(P_{bc}-P_{ab}-1)}^2$, ${(P_{bd}-P_{ab}-1)}^2$ and ${(P_{be}-P_{ab}-1)}^2$.

However, due to constraint (1), only one of the edges $(b,c)$, $(b,d)$ and $(b,e)$ will be part of the cycle and thus have a position number not equal to 0. The other two edges will have position number 0. Thus, these two optimization problems are not satisfied (minimized). For this reason, it would have led to an incorrect result if we had simply inserted all the optimization problems directly into the QUBO matrix, since this would have caused, for example, the term $2P_{ab}^2$ to be inserted into the QUBO matrix three times (in our example), even though only one of these optimization problems can be satisfied. 

Our solution to this problem is simple: we just extracted the part of the optimization problems that concerns only one edge and inserted this part \textit{exactly once} into the QUBO matrix.
\ \\

\begin{algorithm}
  \label{alg:hamcycle}
  \caption{QUBO algorithm for Hamiltonian Cycles}
  \SetKwFunction{algo}{fillQ}
  
  \SetKwProg{myalg}{}{}{}
  \myalg{\algo{$Vertices: V, Edges: E$}}{
  \nl init empty QUBO matrix $Q$\;
  \nl \For{$(a,b) \in E$}{
      \nl \uIf{$b = V_1$}{
        \nl $Q\ \gets2{P_{ab}}^2-2P_{ab}\cdot(\left|V\right|+1)$\;
      }
      \nl \Else{
        \nl $Q\ \gets2{P_{ab}}^2$\;
      }
      \nl \For{$(c,d) \in E$}{
        \nl \uIf{$a = c$ XOR $b = d$}{
            \nl $Q[(a,b),(c,d)] \gets 2{|V|}^2$\;
        }
        \nl \uElseIf{$(b = c \land b \neq V_1) \lor (a = d \land a \neq V_1)$}{
            \nl $Q \gets -2P_{cd}P_{ab}$\;
        }
      }
  }
  \nl \KwRet $Q$\;}{}
  
\end{algorithm}

A brief word on the value $H(x,Q)$: $H$ is usually called ``energy'', following the tradition of physics. The energy for the optimal solution $x^\ast$ for the presented QUBO algorithm for Hamiltonian Cycle is given by $H(x^\ast,Q)=-|V| \cdot\ (|V|+1)$, so we can already infer from the energy value of a solution whether it is indeed correct or not.

\subsection{Size of $Q$ as a function of $|V|$ and $|E|$}

The graph contains at most $2|V|$ edges from or into the start node, which only require one variable each to model the position. For the remaining edges, $\left\lceil{log}_2\left(\left|V\right|+1\right)\right\rceil$ position variables are needed to model the position of each edge. Since there are $|E|$ edges, the size of the QUBO matrix is equal to $n \leq |E| \cdot \lceil{log}_2\left(\left|V\right|+1\right)\rceil$

In the worst case of a fully-connected graph, this QUBO formulation is worse than state-of-the-art. But in settings where the number of edges grows subquadratically as a function of the number of nodes of the graph (as for examples in social networks), the size of $Q$ also grows sub-quadratically. If the number of edges grows linearly in the number of vertices, then this algorithmic QUBO formulation only grows by
$\mathcal{O}(|V| \cdot log(|V|))$ instead of $\mathcal{O}(|V| \cdot |V|)$.

\subsection{Discussion}


To verify the correctness of our QUBO algorithm for Hamiltonian Cycles, we implemented it and applied it to a total of 100 randomly generated graphs with up to 40 nodes and different meshing degrees and solved it using QbSolv~\cite{boothpartitioning}. Since Hamiltonian Cycles, like $k$-SAT, is NP-complete (which means it is also contained in the complexity class NP), it is easy to check whether the solution is indeed correct or not. In all cases, the best QUBO solution corresponded to a valid Hamiltonian cycle which means that the QUBO formulation was indeed correct for these 100 graphs.
\ \\

Since in our QUBO algorithm, unlike in~\cite{16}, the size of the QUBO matrix depends on the number of edges of the graph, we evaluated the scaling behavior for different values of $|E|$ (number of edges) as a function of $|V|$ (number of nodes). Figure 2 shows the scaling behavior for the case of fully-connected graphs.\footnote{We only want to show the scaling behavior here and disregard the meaningfulness, since fully-connected graphs of course trivially contain Hamiltonian Cycles.} The x-axis for the figure describes the number of nodes of the graph and the y-axis represents the size of the resulting QUBO matrix.
\ \\

As can be seen, our algorithm scales significantly worse than the current state-of-the-art formulation in the case of the fully-connected graph.

Figure 3 shows the case where the number of edges grows only linearly in the number of nodes of the graph (in this example $|E|=4|V|$). Here, our method shows a much better scaling behavior than the current method.

\begin{figure}[tb]
 \centering
 \includegraphics[scale=0.55]{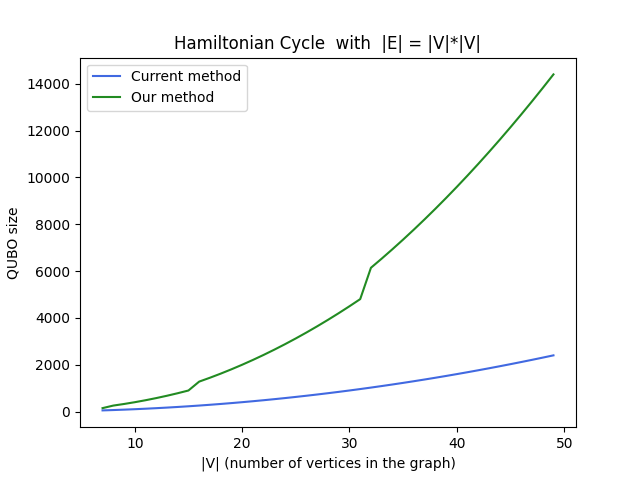}
 \caption{Size of the QUBO matrix as a function of the number of nodes of the graph for the case when the graph is fully-connected.}
\end{figure}

\begin{figure}[tb]
 \centering
 \includegraphics[scale=0.55]{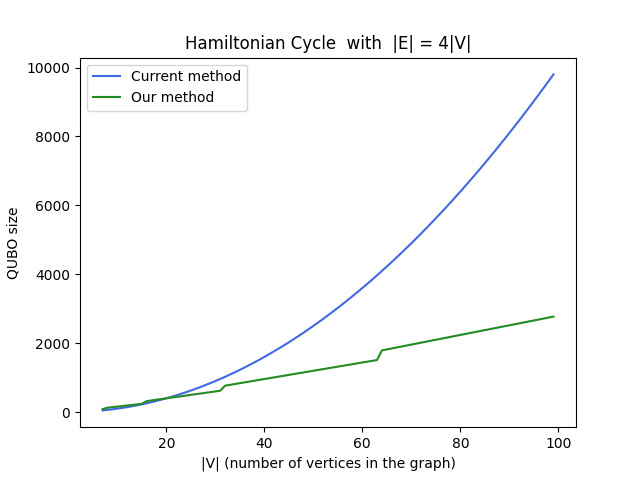}
 \caption{Size of the QUBO matrix as a function of the number of nodes of the graph for the case where the number of edges of the graph grows only linearly in the number of nodes ($|E|=4|V|$).}
\end{figure}

\section{Conclusion}

In this paper, we presented QUBO algorithms (meta-programs) for (Max) $k$-SAT and Hamiltonian Cycles.

The QUBO algorithm for (Max) $k$-SAT only grows logarithmically in $k$, while the current best formulation grows linearly in $k$. We have also shown an algorithmic QUBO formulation for the Hamiltonian Cycle problem that is more efficient if the edges are sufficiently sparse.

To enable larger and more complex QUBO formulations, the devlopment of these formulations must become simpler and more concise. In this paper, we have filled the QUBO piecewise with reusable functions (meta-programs) using the familiar control structures from classical programming, such as loops, branching and recursion.
We have exemplified this method on (Max) $k$-SAT and Hamiltonian Cycles, but we see potential for improvement in numerous other problems in the future.

\ \\
The Python code for the two QUBO algorithms presented here is available on GitHub: \newline
\url{https://github.com/JonasNuesslein/AlgorithmicQUBOs}


\bibliographystyle{ACM-Reference-Format}
\bibliography{main-bib}


\begin{thebibliography}{36}


\ifx \showCODEN    \undefined \def \showCODEN     #1{\unskip}     \fi
\ifx \showDOI      \undefined \def \showDOI       #1{#1}\fi
\ifx \showISBNx    \undefined \def \showISBNx     #1{\unskip}     \fi
\ifx \showISBNxiii \undefined \def \showISBNxiii  #1{\unskip}     \fi
\ifx \showISSN     \undefined \def \showISSN      #1{\unskip}     \fi
\ifx \showLCCN     \undefined \def \showLCCN      #1{\unskip}     \fi
\ifx \shownote     \undefined \def \shownote      #1{#1}          \fi
\ifx \showarticletitle \undefined \def \showarticletitle #1{#1}   \fi
\ifx \showURL      \undefined \def \showURL       {\relax}        \fi
\providecommand\bibfield[2]{#2}
\providecommand\bibinfo[2]{#2}
\providecommand\natexlab[1]{#1}
\providecommand\showeprint[2][]{arXiv:#2}

\bibitem[\protect\citeauthoryear{Albash and Lidar}{Albash and Lidar}{2018}]%
        {albash2018adiabatic}
\bibfield{author}{\bibinfo{person}{Tameem Albash} {and}
  \bibinfo{person}{Daniel~A Lidar}.} \bibinfo{year}{2018}\natexlab{}.
\newblock \showarticletitle{Adiabatic quantum computation}.
\newblock \bibinfo{journal}{\emph{Reviews of Modern Physics}}
  \bibinfo{volume}{90}, \bibinfo{number}{1} (\bibinfo{year}{2018}),
  \bibinfo{pages}{015002}.
\newblock


\bibitem[\protect\citeauthoryear{Applegate, Bixby, Chevátal, and
  Cook}{Applegate et~al\mbox{.}}{2006}]%
        {1}
\bibfield{author}{\bibinfo{person}{David Applegate}, \bibinfo{person}{Robert
  Bixby}, \bibinfo{person}{Vasek Chevátal}, {and} \bibinfo{person}{William
  Cook}.} \bibinfo{year}{2006}\natexlab{}.
\newblock \bibinfo{booktitle}{\emph{The traveling salesman problem: a
  computational study}}.
\newblock


\bibitem[\protect\citeauthoryear{Aramon, Rosenberg, Valiante, Miyazawa, Tamura,
  and Katzgraber}{Aramon et~al\mbox{.}}{2019}]%
        {aramon2019physics}
\bibfield{author}{\bibinfo{person}{Maliheh Aramon}, \bibinfo{person}{Gili
  Rosenberg}, \bibinfo{person}{Elisabetta Valiante}, \bibinfo{person}{Toshiyuki
  Miyazawa}, \bibinfo{person}{Hirotaka Tamura}, {and} \bibinfo{person}{Helmut~G
  Katzgraber}.} \bibinfo{year}{2019}\natexlab{}.
\newblock \showarticletitle{Physics-inspired optimization for quadratic
  unconstrained problems using a digital annealer}.
\newblock \bibinfo{journal}{\emph{Frontiers in Physics}}  \bibinfo{volume}{7}
  (\bibinfo{year}{2019}), \bibinfo{pages}{48}.
\newblock


\bibitem[\protect\citeauthoryear{Barahona}{Barahona}{1982}]%
        {barahona1982computational}
\bibfield{author}{\bibinfo{person}{Francisco Barahona}.}
  \bibinfo{year}{1982}\natexlab{}.
\newblock \showarticletitle{On the computational complexity of Ising spin glass
  models}.
\newblock \bibinfo{journal}{\emph{Journal of Physics A: Mathematical and
  General}} \bibinfo{volume}{15}, \bibinfo{number}{10} (\bibinfo{year}{1982}),
  \bibinfo{pages}{3241}.
\newblock


\bibitem[\protect\citeauthoryear{Bian, Chudak, Macready, Roy, Sebastiani, and
  Varotti}{Bian et~al\mbox{.}}{2018}]%
        {7}
\bibfield{author}{\bibinfo{person}{Zhengbing Bian}, \bibinfo{person}{Fabian
  Chudak}, \bibinfo{person}{William Macready}, \bibinfo{person}{Aidan Roy},
  \bibinfo{person}{Roberto Sebastiani}, {and} \bibinfo{person}{Stefano
  Varotti}.} \bibinfo{year}{2018}\natexlab{}.
\newblock \showarticletitle{Solving {SAT} and {MaxSAT} with a Quantum Annealer:
  Foundations, Encodings, and Preliminary Results}.
\newblock  (\bibinfo{year}{2018}).
\newblock


\bibitem[\protect\citeauthoryear{Blum and Rivest}{Blum and Rivest}{1988}]%
        {3}
\bibfield{author}{\bibinfo{person}{Avrim Blum} {and} \bibinfo{person}{Ronald
  Rivest}.} \bibinfo{year}{1988}\natexlab{}.
\newblock \showarticletitle{Training a 3-node neural network is {NP}-complete}.
\newblock  (\bibinfo{year}{1988}).
\newblock


\bibitem[\protect\citeauthoryear{Booth, Reinhardt, and Roy}{Booth
  et~al\mbox{.}}{[n.\,d.]}]%
        {boothpartitioning}
\bibfield{author}{\bibinfo{person}{M Booth}, \bibinfo{person}{SP Reinhardt},
  {and} \bibinfo{person}{A Roy}.} \bibinfo{year}{[n.\,d.]}\natexlab{}.
\newblock \bibinfo{title}{Partitioning optimization problems for hybrid
  classica/quantum execution (2017)}.
\newblock
\newblock


\bibitem[\protect\citeauthoryear{Chancellor, Zohren, Warburton, Benjamin, and
  Roberts}{Chancellor et~al\mbox{.}}{2016}]%
        {14}
\bibfield{author}{\bibinfo{person}{Nick Chancellor}, \bibinfo{person}{S
  Zohren}, \bibinfo{person}{P~A Warburton}, \bibinfo{person}{S~C Benjamin},
  {and} \bibinfo{person}{S Roberts}.} \bibinfo{year}{2016}\natexlab{}.
\newblock \showarticletitle{A Direct Mapping of Max k-{SAT} and High Order
  Parity Checks to a Chimera Graph}.
\newblock  (\bibinfo{year}{2016}).
\newblock


\bibitem[\protect\citeauthoryear{Choi}{Choi}{2010}]%
        {13}
\bibfield{author}{\bibinfo{person}{Vicky Choi}.}
  \bibinfo{year}{2010}\natexlab{}.
\newblock \showarticletitle{Adiabatic Quantum Algorithms for the {NP}-Complete
  Maximum-Weight Independent Set, Exact Cover and {3SAT} Problems}.
\newblock  (\bibinfo{year}{2010}).
\newblock


\bibitem[\protect\citeauthoryear{Choi}{Choi}{2011}]%
        {8}
\bibfield{author}{\bibinfo{person}{Vicky Choi}.}
  \bibinfo{year}{2011}\natexlab{}.
\newblock \showarticletitle{Different Adiabatic Quantum Optimization Algorithms
  for the {NP}-Complete Exact Cover and {3SAT} Problems}.
\newblock  (\bibinfo{year}{2011}).
\newblock


\bibitem[\protect\citeauthoryear{Cook}{Cook}{2000}]%
        {cook2000p}
\bibfield{author}{\bibinfo{person}{Stephen Cook}.}
  \bibinfo{year}{2000}\natexlab{}.
\newblock \showarticletitle{The P versus NP problem}.
\newblock \bibinfo{journal}{\emph{Clay Mathematics Institute}}
  \bibinfo{volume}{2} (\bibinfo{year}{2000}).
\newblock


\bibitem[\protect\citeauthoryear{Dasgupta}{Dasgupta}{2008}]%
        {2}
\bibfield{author}{\bibinfo{person}{Sanjoy Dasgupta}.}
  \bibinfo{year}{2008}\natexlab{}.
\newblock \bibinfo{booktitle}{\emph{The Hardness of K-means Clustering}}.
\newblock


\bibitem[\protect\citeauthoryear{E{\'e}n and S{\"o}rensson}{E{\'e}n and
  S{\"o}rensson}{2003}]%
        {een2003extensible}
\bibfield{author}{\bibinfo{person}{Niklas E{\'e}n} {and}
  \bibinfo{person}{Niklas S{\"o}rensson}.} \bibinfo{year}{2003}\natexlab{}.
\newblock \showarticletitle{An extensible SAT-solver}. In
  \bibinfo{booktitle}{\emph{International conference on theory and applications
  of satisfiability testing}}. Springer, \bibinfo{pages}{502--518}.
\newblock


\bibitem[\protect\citeauthoryear{Farhi, Goldstone, and Gutmann}{Farhi
  et~al\mbox{.}}{2014}]%
        {farhi2014quantum}
\bibfield{author}{\bibinfo{person}{Edward Farhi}, \bibinfo{person}{Jeffrey
  Goldstone}, {and} \bibinfo{person}{Sam Gutmann}.}
  \bibinfo{year}{2014}\natexlab{}.
\newblock \showarticletitle{A quantum approximate optimization algorithm}.
\newblock \bibinfo{journal}{\emph{arXiv preprint arXiv:1411.4028}}
  (\bibinfo{year}{2014}).
\newblock


\bibitem[\protect\citeauthoryear{Finnila, Gomez, Sebenik, Stenson, and
  Doll}{Finnila et~al\mbox{.}}{1994}]%
        {finnila1994quantum}
\bibfield{author}{\bibinfo{person}{Aleta~Berk Finnila}, \bibinfo{person}{MA
  Gomez}, \bibinfo{person}{C Sebenik}, \bibinfo{person}{Catherine Stenson},
  {and} \bibinfo{person}{Jimmie~D Doll}.} \bibinfo{year}{1994}\natexlab{}.
\newblock \showarticletitle{Quantum annealing: A new method for minimizing
  multidimensional functions}.
\newblock \bibinfo{journal}{\emph{Chemical physics letters}}
  \bibinfo{volume}{219}, \bibinfo{number}{5-6} (\bibinfo{year}{1994}),
  \bibinfo{pages}{343--348}.
\newblock


\bibitem[\protect\citeauthoryear{Glover, Kochenberger, and Du}{Glover
  et~al\mbox{.}}{2018}]%
        {glover2018tutorial}
\bibfield{author}{\bibinfo{person}{Fred Glover}, \bibinfo{person}{Gary
  Kochenberger}, {and} \bibinfo{person}{Yu Du}.}
  \bibinfo{year}{2018}\natexlab{}.
\newblock \showarticletitle{A tutorial on formulating and using QUBO models}.
\newblock \bibinfo{journal}{\emph{arXiv preprint arXiv:1811.11538}}
  (\bibinfo{year}{2018}).
\newblock


\bibitem[\protect\citeauthoryear{Glover, Kochenberger, and Du}{Glover
  et~al\mbox{.}}{2019}]%
        {15}
\bibfield{author}{\bibinfo{person}{Fred Glover}, \bibinfo{person}{Gary
  Kochenberger}, {and} \bibinfo{person}{Yu Du}.}
  \bibinfo{year}{2019}\natexlab{}.
\newblock \showarticletitle{Quantum Bridge Analytics {I}: A Tutorial on
  Formulating and Using {QUBO} Models}.
\newblock  (\bibinfo{year}{2019}).
\newblock


\bibitem[\protect\citeauthoryear{Grover}{Grover}{1996a}]%
        {29}
\bibfield{author}{\bibinfo{person}{Lov Grover}.}
  \bibinfo{year}{1996}\natexlab{a}.
\newblock \showarticletitle{A fast quantum mechanical algorithm for database
  search}.
\newblock  (\bibinfo{year}{1996}).
\newblock


\bibitem[\protect\citeauthoryear{Grover}{Grover}{1996b}]%
        {grover1996fast}
\bibfield{author}{\bibinfo{person}{Lov~K Grover}.}
  \bibinfo{year}{1996}\natexlab{b}.
\newblock \showarticletitle{A fast quantum mechanical algorithm for database
  search}. In \bibinfo{booktitle}{\emph{Proceedings of the twenty-eighth annual
  ACM symposium on Theory of computing}}. \bibinfo{pages}{212--219}.
\newblock


\bibitem[\protect\citeauthoryear{Impagliazzo and Paturi}{Impagliazzo and
  Paturi}{1999}]%
        {20}
\bibfield{author}{\bibinfo{person}{R Impagliazzo} {and} \bibinfo{person}{R
  Paturi}.} \bibinfo{year}{1999}\natexlab{}.
\newblock \showarticletitle{Complexity of k-{SAT}}.
\newblock  (\bibinfo{year}{1999}).
\newblock


\bibitem[\protect\citeauthoryear{Impagliazzo and Paturi}{Impagliazzo and
  Paturi}{2001}]%
        {impagliazzo2001complexity}
\bibfield{author}{\bibinfo{person}{Russell Impagliazzo} {and}
  \bibinfo{person}{Ramamohan Paturi}.} \bibinfo{year}{2001}\natexlab{}.
\newblock \showarticletitle{On the complexity of k-SAT}.
\newblock \bibinfo{journal}{\emph{J. Comput. System Sci.}}
  \bibinfo{volume}{62}, \bibinfo{number}{2} (\bibinfo{year}{2001}),
  \bibinfo{pages}{367--375}.
\newblock


\bibitem[\protect\citeauthoryear{Kaminsky and Lloyd}{Kaminsky and
  Lloyd}{2004}]%
        {kaminsky2004scalable}
\bibfield{author}{\bibinfo{person}{William~M Kaminsky} {and}
  \bibinfo{person}{Seth Lloyd}.} \bibinfo{year}{2004}\natexlab{}.
\newblock \showarticletitle{Scalable architecture for adiabatic quantum
  computing of NP-hard problems}.
\newblock \bibinfo{journal}{\emph{Quantum computing and quantum bits in
  mesoscopic systems}} (\bibinfo{year}{2004}), \bibinfo{pages}{229--236}.
\newblock


\bibitem[\protect\citeauthoryear{Karp}{Karp}{1972}]%
        {22}
\bibfield{author}{\bibinfo{person}{Richard~Manning Karp}.}
  \bibinfo{year}{1972}\natexlab{}.
\newblock \showarticletitle{Reducibility Among Combinatorial Problems}.
\newblock  (\bibinfo{year}{1972}).
\newblock


\bibitem[\protect\citeauthoryear{Krom}{Krom}{1967}]%
        {21}
\bibfield{author}{\bibinfo{person}{M~R Krom}.} \bibinfo{year}{1967}\natexlab{}.
\newblock \showarticletitle{The Decision Problem for a Class of First‐Order
  Formulas in Which all Disjunctions are Binary}.
\newblock  (\bibinfo{year}{1967}).
\newblock


\bibitem[\protect\citeauthoryear{Lodewijks}{Lodewijks}{2020}]%
        {12}
\bibfield{author}{\bibinfo{person}{Bas Lodewijks}.}
  \bibinfo{year}{2020}\natexlab{}.
\newblock \showarticletitle{Mapping {NP}-hard and {NP}-complete Optimisation
  Problems to Quadratic Unconstrained Binary Optimisation Problems}.
\newblock  (\bibinfo{year}{2020}).
\newblock


\bibitem[\protect\citeauthoryear{Lucas}{Lucas}{2014}]%
        {16}
\bibfield{author}{\bibinfo{person}{Andrew Lucas}.}
  \bibinfo{year}{2014}\natexlab{}.
\newblock \showarticletitle{{Ising} formulations of many {NP} problems}.
\newblock  (\bibinfo{year}{2014}).
\newblock


\bibitem[\protect\citeauthoryear{Mahasinghe, Hua, Dinneen, and
  Goyal}{Mahasinghe et~al\mbox{.}}{2019}]%
        {27}
\bibfield{author}{\bibinfo{person}{Anuradha Mahasinghe},
  \bibinfo{person}{Richard Hua}, \bibinfo{person}{Michael Dinneen}, {and}
  \bibinfo{person}{Rajni Goyal}.} \bibinfo{year}{2019}\natexlab{}.
\newblock \showarticletitle{Solving the Hamiltonian Cycle Problem using a
  Quantum Computer}.
\newblock  (\bibinfo{year}{2019}).
\newblock


\bibitem[\protect\citeauthoryear{McGeoch}{McGeoch}{2014}]%
        {mcgeoch2014adiabatic}
\bibfield{author}{\bibinfo{person}{Catherine~C McGeoch}.}
  \bibinfo{year}{2014}\natexlab{}.
\newblock \showarticletitle{Adiabatic quantum computation and quantum
  annealing: Theory and practice}.
\newblock \bibinfo{journal}{\emph{Synthesis Lectures on Quantum Computing}}
  \bibinfo{volume}{5}, \bibinfo{number}{2} (\bibinfo{year}{2014}),
  \bibinfo{pages}{1--93}.
\newblock


\bibitem[\protect\citeauthoryear{Mooney, Tonetto, Hill, and Hollenberg}{Mooney
  et~al\mbox{.}}{2019}]%
        {10}
\bibfield{author}{\bibinfo{person}{Gary Mooney}, \bibinfo{person}{Sam Tonetto},
  \bibinfo{person}{Charles Hill}, {and} \bibinfo{person}{Lloyd Hollenberg}.}
  \bibinfo{year}{2019}\natexlab{}.
\newblock \showarticletitle{Mapping {NP}-hard Problems to restructed Adiabatic
  Quantum Architectures}.
\newblock  (\bibinfo{year}{2019}).
\newblock


\bibitem[\protect\citeauthoryear{Pakin}{Pakin}{2018}]%
        {30}
\bibfield{author}{\bibinfo{person}{Scott Pakin}.}
  \bibinfo{year}{2018}\natexlab{}.
\newblock \showarticletitle{Performing Fully Parallel Constraint Logic
  Programming on a Quantum Annealer}.
\newblock  (\bibinfo{year}{2018}).
\newblock


\bibitem[\protect\citeauthoryear{Prasanna, Patton, Schuman, and Potok}{Prasanna
  et~al\mbox{.}}{2019}]%
        {4}
\bibfield{author}{\bibinfo{person}{Date Prasanna}, \bibinfo{person}{Robert
  Patton}, \bibinfo{person}{Catherine Schuman}, {and} \bibinfo{person}{Thomas
  Potok}.} \bibinfo{year}{2019}\natexlab{}.
\newblock \showarticletitle{Efficiently embedding {QUBO} problems on adiabatic
  quantum computers}.
\newblock  (\bibinfo{year}{2019}).
\newblock


\bibitem[\protect\citeauthoryear{Proctor, Rudinger, Young, Nielsen, and
  Blume-Kohout}{Proctor et~al\mbox{.}}{2022}]%
        {proctor2022measuring}
\bibfield{author}{\bibinfo{person}{Timothy Proctor}, \bibinfo{person}{Kenneth
  Rudinger}, \bibinfo{person}{Kevin Young}, \bibinfo{person}{Erik Nielsen},
  {and} \bibinfo{person}{Robin Blume-Kohout}.} \bibinfo{year}{2022}\natexlab{}.
\newblock \showarticletitle{Measuring the capabilities of quantum computers}.
\newblock \bibinfo{journal}{\emph{Nature Physics}} \bibinfo{volume}{18},
  \bibinfo{number}{1} (\bibinfo{year}{2022}), \bibinfo{pages}{75--79}.
\newblock


\bibitem[\protect\citeauthoryear{Rudolph}{Rudolph}{1995}]%
        {26}
\bibfield{author}{\bibinfo{person}{T. Rudolph}.}
  \bibinfo{year}{1995}\natexlab{}.
\newblock \showarticletitle{Quantum Computing Hamiltonian cycles.}
\newblock  (\bibinfo{year}{1995}).
\newblock


\bibitem[\protect\citeauthoryear{Shor}{Shor}{1999}]%
        {shor1999polynomial}
\bibfield{author}{\bibinfo{person}{Peter~W Shor}.}
  \bibinfo{year}{1999}\natexlab{}.
\newblock \showarticletitle{Polynomial-time algorithms for prime factorization
  and discrete logarithms on a quantum computer}.
\newblock \bibinfo{journal}{\emph{SIAM review}} \bibinfo{volume}{41},
  \bibinfo{number}{2} (\bibinfo{year}{1999}), \bibinfo{pages}{303--332}.
\newblock


\bibitem[\protect\citeauthoryear{Vargas-Calderón, Parra-A., and
  Vinck-Posada}{Vargas-Calderón et~al\mbox{.}}{2021}]%
        {23}
\bibfield{author}{\bibinfo{person}{Vladimir Vargas-Calderón},
  \bibinfo{person}{Nicolas Parra-A.}, {and} \bibinfo{person}{Herbert
  Vinck-Posada}.} \bibinfo{year}{2021}\natexlab{}.
\newblock \showarticletitle{Many-Qudit representation for the Travelling
  Salesman Problem Optimisation}.
\newblock  (\bibinfo{year}{2021}).
\newblock


\bibitem[\protect\citeauthoryear{Zahedinejad and Zaribafiyan}{Zahedinejad and
  Zaribafiyan}{2017}]%
        {zahedinejad2017combinatorial}
\bibfield{author}{\bibinfo{person}{Ehsan Zahedinejad} {and}
  \bibinfo{person}{Arman Zaribafiyan}.} \bibinfo{year}{2017}\natexlab{}.
\newblock \showarticletitle{Combinatorial optimization on gate model quantum
  computers: A survey}.
\newblock \bibinfo{journal}{\emph{arXiv preprint arXiv:1708.05294}}
  (\bibinfo{year}{2017}).
\newblock


\end{thebibliography}

\end{document}